\newcommand{\be}{\begin{equation}}\newcommand{\ee}{\end{equation}}
\newcommand{\bea}{\begin{eqnarray}}\newcommand{\eea}{\end{eqnarray}}
\newcommand{\nn}{\nonumber}\newcommand{\p}[1]{(\ref{#1})}
\newcommand{\lb}{\label}

\newcommand\s{\scriptscriptstyle}
\newcommand{\ab}{{\alpha\beta}}
\newcommand{\pab}{\partial_\ab}
\newcommand{\ta}{\theta^\alpha}
\newcommand{\tb}{\theta^\beta}
\newcommand{\bta}{\bar{\theta}^\alpha}
\newcommand{\btb}{\bar{\theta}^\beta}
\newcommand{\tao}{\theta^\alpha_1}

\newcommand{\tat}{\theta^\alpha_2}
\newcommand{\ts}{(\theta)^2}
\newcommand{\bts}{(\bar{\theta})^2}
\newcommand{\tbt}{(\theta\bar{\theta})}
\newcommand{\too}{(\theta_1\theta_1)}
\newcommand{\ttt}{(\theta_2\theta_2)}
\newcommand\cD{{\cal D}}
\newcommand\bcD{\bar{\cal D}}
\newcommand{\cDa}{\cD_\alpha}

\newcommand{\bcDa}{\bcD_\alpha}

\newcommand\cDao{{\cal D}_\alpha^1}
\newcommand\cDat{{\cal D}_\alpha^2}

\newcommand{\Da}{D_\alpha}

\newcommand{\bDa}{\bar{D}_\alpha}

\newcommand{\Ds}{(D)^2}
\newcommand{\bDs}{(\bar{D})^2}
\newcommand{\DbD}{(D\bar{D})}
\newcommand{\Dao}{D_\alpha^1}
\newcommand{\Dat}{D_\alpha^2}

\newcommand{\Doo}{(D^1D^1)}

\documentstyle[12pt]{article}

\begin{document}
\begin{flushright}
{ hep-th/9911166 }
\end{flushright}
\vspace{2cm}

\begin{center}
{\large\bf GOLDSTONE-TYPE SUPERFIELDS AND
PARTIAL SPONTANEOUS BREAKING OF D=3, N=2 SUPERSYMMETRY}

\vspace{0.5cm}
{\bf  B.M. Zupnik}\footnote{On leave on absence from the Institute
of Applied Physics, Tashkent State University, Uzbekistan\\
E-mail: zupnik@thsun1.jinr.ru}\\
{\it Bogoliubov Laboratory of Theoretical Physics, Joint Institute
for Nuclear Research, 141980, Dubna, Russia}\\

\end{center}

\begin{abstract}
We consider the modified superfield constraints with constant terms for
the $D{=}3$, $~N{=}2$ Goldstone-Maxwell gauge multiplet which contains
Goldstone fermions , real scalar and vector fields. The partial
spontaneous breaking  $N{=}2\rightarrow N{=}1$ is possible for the
non-minimal self-interaction of this modified gauge superfield including
the linear Fayet-Iliopoulos term. The dual description of the partial
breaking in the model of the self-interacting Goldstone chiral superfield
is also discussed.
\end{abstract}



\setcounter{footnote}0

\setcounter{equation}0
\section{\lb{B}Goldstone-Maxwell superfield }

  Models with the partial spontaneous breaking of the global
$D{=}3,~N{=}2$  supersymmetry  have been constructed using the
topologically non-trivial classical solutions preserving the one half
of supercharges \cite{AGIT} and also  in the method of nonlinear
realizations of supersymmetries using  superfields of the unbroken
$N{=}1$ supersymmetry \cite{IK}. These models describe interactions of
the Goldstone fermions with the complex scalar field of the supermembrane
or with the real scalar and vector fields of the $D2$-brane.

The standard linear supermultiplets (standard superfields) are not
convenient for a description of the  partial spontaneous  breaking
of the extended global supersymmetries $(PSBGS)$ when the invariance with
respect to the part of supercharges remains  unbroken. The partial
breaking of the $N{=}2$ supersymmetry means a degeneracy of the matrix
of vacuum supersymmetric transformations of two real fermions in some
supermultiplet , however, this is impossible for the standard structure of
auxiliary components in the vector or chiral superfields. Nevertheless,
the linear $N{=}2$ transformations with the partial breaking can be
constructed in terms of the $N{=}1$ superfields \cite{IK}. We shall show
that these representations with the linear Goldstone ($LG$) fermions
correspond to the modified constraints for the Goldstone-type superfields
in the $N{=}2$ superspace.

The Goldstone-Maxwell chiral superfield $W$ in the  $D{=}4,~N{=}2$
superspace satisfies the modified superfield 2-nd order constraints
\cite{APT,IZ}. In comparison to the original constraints of the $N{=}2$
vector multiplet \cite{GSW}, the deformed constraints contain the {\it
constant} terms which guarantee the appearance of the unusual constant
imaginary part of the isovector auxiliary component and the  Goldstone
fermion component in the gauge superfield. This abelian gauge model
has been used to break spontaneously $D{=}4,~N{=}2$ supersymmetry to its
$N{=}1$ subgroup.

The more early example of the Goldstone-type constraint has been
considered in the model with the partial breaking of the $D{=}1,~N{=}4$
supersymmetry \cite{IKP}. Thus, these constraints introduce a new type of
the supersymmetry representations with the  $LG$-fermions. In distinction
with the Goldstone fermions of the nonlinear realizations which transforms
linearly only in the unbroken supersymmetry, the $LG$-fermions have their
partners in the  supermultiplets of the whole supersymmetry. The nonlinear
deformation of the standard constraints is also possible \cite{GGRS},
however, we shall discuss only constant terms in the modified constraints
which are connected with the spontaneous breaking of supersymmetries. It
will be shown that the models with the $LG$ vector multiplet and the
corresponding dual scalar multiplet solve the problem of the partial
spontaneous breaking of the $D{=}3,~N{=}2$ supersymmetry.

In this section we discuss the prepotential solution for the $LGM$
supermultiplet which contains additional terms manifestly depending on the
spinor coordinates and some complex constants  playing the role of moduli
in the vacuum state of the theory together with the constant of the
Fayet-Iliopoulos $(FI)$ term. Using this representation in the non-minimal
gauge action one can obtain the constant vacuum solutions with the partial
spontaneous breaking of the $D{=}3,~N{=}2$ supersymmetry. Note that the
supersymmetry algebra is modified on the $LGM$ prepotential $V$ by analogy
with the similar modified transformations of the $4D$ gauge fields or
prepotentials in refs.\cite{IZ,FGP}.

The sect.\ref{C} is devoted to the description of  $PSBGS$ in the
interaction of the $LG$-chiral superfield which is dual to the interaction
of the $LGM$  superfield. This manifestly supersymmetric action depends on
the sum of the chiral and antichiral superfields and some constant term
bilinear in the spinor coordinates. The non-usual transformation of the
basic $LG$-chiral superfield satisfies the supersymmetry algebra with the
central-charge term.

The $N{=}1$ supermembrane and $D2$-brane actions \cite{IK} can be analysed
in our approach using the decompositions of $N{=}2$ superfields in the
2-nd spinor coordinate $\theta^\alpha_2$. We consider the $N{=}1$
components of the extended superfields and the covariant conditions which
allow us to express the additional degrees of freedom in terms of the
Goldstone superfields.

The coordinates of the full $D{=}3,~N{=}2$ superspace are
\be
z=(x^\ab,\ta ,\bta )~,\lb{B1}
\ee
where $\alpha,\beta$ are the spinor indices of the group $SL(2,R)$.
The spinor representation of the coordinate is connected with the
vector representation via the $3D$ $\gamma$-matrices $x^\ab{=}(1/2)x^m
(\gamma_m)^\ab$. The  spinor derivatives in this superspace have
the following form:
\bea
&&\cDa=\Da + {i\over2}\bar{\theta}_\alpha Z~,\qquad \Da =\partial_\alpha
+{i\over2}\btb\pab~,\nn\\
&&\bcDa=\bDa -{i\over2}\theta_\alpha Z~,\qquad \bDa =\bar{\partial}_\alpha
+{i\over2}\tb\pab,\lb{B4}
\eea
where $Z$ is the real central charge, and $\Da$ and $\bDa$ are
 the spinor derivatives without the central charge.

The $N{=}2$ supersymmetry algebra  is covariant with respect to the
$U_R(1)$ transformations of the spinor coordinates.

 We shall consider the following notation for the bilinear combinations of
spinor coordinates and differential operators:
\bea
&&\ts={1\over2}\theta_\alpha \theta^\alpha~,\quad \bts =
{1\over2} \bar{\theta}^\alpha \bar{\theta}_\alpha~,
\lb{B5}\\
&&\tbt ={1\over2}\ta\bar{\theta}_\alpha~,\qquad\Theta^{\alpha\beta}
={1\over2}  [\theta^\alpha\bar{\theta}^\beta +\alpha  \leftrightarrow
\beta ]~,\lb{B6}\\
&& \Ds ={1\over2}D^\alpha D_\alpha~,\qquad \bDs ={1\over2}
   \bar{D}_\alpha \bar{D}^\alpha~,\lb{B7}\\
   &&\DbD ={1\over2}D^\alpha\bar{D}_\alpha~, \qquad D_{\alpha\beta}=
{1\over2}([D_\alpha,\bar{D}_\beta]+ \alpha  \leftrightarrow\beta)~.
\lb{B8}
\eea

 The complex chiral coordinates can be constructed by the analogy with
$D{=}4$
\be
\zeta=(x_{\s L}^\ab,\ta)~,\qquad x_{\s L}^\ab=x^\ab+i\Theta^\ab~.\lb{B10}
\ee

It is convenient to use the following rules of conjugation for any
operators
\cite{GGRS}:
\be
(X Y)^\dagger=Y^\dagger X^\dagger~,\qquad [X,Y\}^\dagger=-(-1)^{p(X)p(Y)}
[X^\dagger,Y^\dagger\}~,\lb{B11}
\ee
where $[X,Y\}$ is the graded commutator and $p(X)=\pm1$ is the
$Z_2$-parity.

It is possible to introduce the real $N{=}2$ spinor coordinates
$\theta^\alpha_k=(\theta^\alpha_k)^\dagger $
\bea
&&\tao={1\over\sqrt{2}}(\ta +\bta )~,
\qquad \tat ={i\over\sqrt{2}}(\bta -\ta )~,\lb{B12}\\
&& (\theta\bar{\theta})={1\over2}[ (\theta_1\theta_1)+
(\theta_2\theta_2)]~,\qquad (\theta_i\theta_k)^\dagger=
-(\theta_i\theta_k)\lb{B14}
\eea
and the corresponding real spinor derivatives
\bea
&&\cDao=\Dao +{1\over2}\theta_{2\alpha}Z~,\qquad \cDat=\Dat -{1\over2}
\theta_{1\alpha}Z~,\lb{B15}\\
&& D^1_\alpha={1\over\sqrt{2}}(D_\alpha +\bar{D}_\alpha)~,\qquad
D^2_\alpha={i\over\sqrt{2}}(D_\alpha -\bar{D}_\alpha)~.\lb{B16}
\eea

The  $D{=}3,~ N{=}2$ gauge theory \cite{Si,ZP1,BHO} is analogous to the
well-known $D{=}4,~ N{=}1$ gauge theory, although the three-dimensional
case has some interesting peculiarities which are connected with the
existence of the topological mass term and duality between the $3D$-vector
and chiral multiplets. We shall consider the basic superspace with
$Z{=}0$.

The abelian $U(1)$-gauge prepotential $V(z)$  possesses the gauge
transformation
$\delta V=\Lambda +\bar{\Lambda}$
where  $\Lambda$ is the chiral
parameter.

The $D{=}3,~ N{=}2$ vector multiplet is described by the real linear
superfield
\be
W(V)= i(D\bar{D})V\lb{B19}
\ee
satisfying the basic constraints $(D)^2W=(\bar{D})^2W=0$.

The  components of $W(V)$ are  the real scalar $\varphi$, the
field-strength of the gauge field $F_{\alpha\beta}(A)$, the real
auxiliary component $G$ and the spinor fields $ \lambda^\alpha$ and
$\bar{\lambda}^\alpha$.

The low-energy effective action of the $3D$ vector multiplet describes
a non-minimal interaction of the real scalar field with the fermion and
gauge fields. For the $U(1)$  gauge superfield $V$ this action has the
following general form:
\be
S(W)=-{1\over2}\int d^7z H(W)~,\qquad \tau(W)=H^{\prime\prime}(W) > 0~,
\lb{B20}
\ee
where $H(W) $ is the real convex function of $W$.

Let us consider the spontaneous breaking of supersymmetry in the
non-minimal gauge model (\ref{B20}) with the additional linear $FI$-term
\be
S_{\s FI}={1\over2} \xi\int d^7z V~,\lb{B21}
\ee
where $\xi$ is a constant of the dimension $-1$.  Varying the superfield
$V$ one can derive the corresponding superfield equation of motion. We
shall study the constant solutions of this equation  using the following
vacuum ansatz:
\be
V_0=2i\tbt a -2\ts\bts G~,\qquad W_0=a + 2i\tbt G~,\lb{B22}
\ee
where $a$ and $G$ are constants. The non-trivial solution $G_0{\neq}0$ is
possible for the quadratic function $H$ only.

It is useful to consider the real  spinors $\lambda^\alpha_k$ and the real
spinor parameters of the $N{=}2$ supersymmetry by analogy with \p{B12}.
It is clear that the constant solution $G_0{\neq}0$ can only break
spontaneously {\it both} supersymmetries.

Consider the following  deformation of the linearity constraints which
defines the  $LGM$ superfield:
\be
(D)^2\hat{W}=C~,\qquad (\bar{D})^2\hat{W}=\bar{C}~,\lb{B27}
\ee
where $C$ and $\bar{C}$ are some constants. These relations are manifestly
supersymmetric, however, they break the $U_R(1)$ invariance.

The gauge-prepotential solution of these constraints can be constructed
by analogy with Eq.(\ref{B19})
\be
\hat{W}=i\DbD V+(\theta)^2C
+(\bar{\theta})^2\bar{C}~.\lb{B28}
\ee
This  superfield contains new constant auxiliary structures which
change radically the matrix of the vacuum fermion transformations
$\delta_\epsilon\lambda^\alpha_k$. It is evident that the
$PSBGS$-condition corresponds to the degeneracy of these transformations
\be
C\bar{C}-G^2_0=0~.\lb{B31}
\ee
In this case one can choose the single real Goldstone spinor field as
some linear combination of $\lambda^\alpha_k$.

The action of the $LGM$-superfield (\ref{B28}) has the following form:
\be
\hat{S}(V)=-{1\over2}\int d^7z [H(\hat{W})-\xi V]\lb{B32}
\ee
and depends on three constants $\xi, C$ and $\bar{C}$.

The non-derivative terms in the component Lagrangian
produce the following scalar potential
\be
{\cal V}(\varphi)={1\over2}[|C|^2\tau(\varphi)+\xi^2\tau^{-1}(\varphi)]~.
\lb{B34}
\ee

The $PSBGS$ solution (\ref{B31}) arises for the non-trivial interaction
$\tau^\prime(a){\neq}0$. This solution determines the minimum point
$a_{\s0}$ of this model
\be
\tau(a_{\s0})=\frac{|\xi|}{|C|}~.\lb{B35}
\ee

The vacuum auxiliary field can be calculated in the point $a_{\s0}$
\be
G_{\s0}=\frac{\xi}{\tau(a_{\s0})}=\pm|C|~.\lb{B36}
\ee
Using the $U_R(1)$ transformation one can choose the pure imaginary
constant $C\rightarrow c=i|c|$ (without the loss of generality) then
$G_{\s0}=-ic=|c|$.

This choice corresponds to the following decomposition of the
$LGM$-superfield (\ref{B28})
\bea
&&\hat{W}=W_s(V_s) +2i|c|\ttt~,\lb{B38}\\
&& W_s(V_s)={i\over4}(D^{1\alpha}D^1_\alpha+D^{2\alpha}D^2_\alpha)
V_s~.\lb{B39}
\eea
where $V_s$ is the shifted $LGM$-prepotential which has  the vanishing
vacuum solution for the auxiliary component. It is evident that this
representation breaks spontaneously the 2-nd supersymmetry only.

It should be stressed that the shifted  quantities $W_s$ and $V_s$ are not
standard superfields
\bea
&&\delta_\epsilon W_s=i\epsilon^\alpha_kQ_\alpha^k \hat{W}=
-2i|c|\epsilon^\alpha_2\theta_{2\alpha}+ i\epsilon^\alpha_k
Q_\alpha^k W_s~,
\lb{B40}\\
&& \delta_\epsilon V_s=2|c|\epsilon^\alpha_2\theta_{2\alpha}
(\theta^\beta_1\theta_{1\beta}) +
i\epsilon^\alpha_k Q_\alpha^k V_s
~.\lb{B41}
\eea

The supersymmetry algebra of the $V_s$-transformations is essentially
modified by the analogy with the transformations of the prepotentials
in refs.\cite{IZ,FGP}.

It should be remarked that the minimal interaction of the charged chiral
superfields with the $LGM$-prepotential $V_s$ breaks the supersymmetry.
The analogous problem of the $LGM$ interaction with the charged matter
appears also in the $PSBGS$ model with $D{=}4,~N{=}2$ supersymmetry
\cite{IZ}.

\setcounter{equation}0
\section{\lb{C}Goldstone chiral superfield}

The $3D$ linear multiplet is dual to the chiral multiplet $\phi$.
The Legendre transform describing  this duality is
\be
S[B,\Phi]=-{1\over2}\int d^7z [H(B)-\Phi B]~,\lb{C1}
\ee
where $B$ is the real unconstrained superfield and $\Phi{=}\phi +
\bar{\phi}$. Varying the Lagrange multipliers $\phi$ and $\bar{\phi}$
one can obtain the linearity constraints for $B$.

We shall show that the spontaneous breaking of supersymmetry is possible
for the non-trivial interaction of the $LG$ chiral superfield which
possesses the inhomogeneous supersymmetry transformation. Let us consider
the dual picture for the $PSBGS$ gauge model with the $FI$-term \p{B32}
\be
\hat{S}(B,\phi,\bar{\phi})=-{1\over2}\int d^7z [H(B) - B\hat{\Phi}]
-{1\over2}[\bar{C}\int d^3x(D)^2 \phi +\mbox{c.c.}]~,
\lb{C4}
\ee
where the modified constrained $LG$ superfield is introduced
\be
\hat{\Phi}\equiv \phi+\bar{\phi}+ 2i\xi\tbt~,\qquad \DbD\hat{\Phi}=-i\xi~.
\lb{C5}
\ee

Varying the chiral and antichiral Lagrange multipliers $\phi$ and
$\bar{\phi}$ one can obtain the $LGM$-constraints (\ref{B27}) on the
superfield $B$ and then pass to the gauge phase $B\rightarrow \hat{W}(V)$
where the $\tbt$-term in $\hat{\Phi}$ transforms to the $FI$-term.

The algebraic $B$-equation
\be
H^\prime(B)\equiv f(B)=\hat{\Phi}
~,\lb{C6}
\ee
provides the transform to the `chiral' phase
\be
B~\rightarrow~f^{-1}(\hat{\Phi})\equiv\hat{B}(\hat{\Phi})~.\lb{C7}
\ee

The transformed chiral action is
\bea
&&\hat{S}(\hat{\Phi})=-{1\over2}\int d^7z  \{\hat{H}(\hat{\Phi})
+[\bar{C}\ts +\mbox{c.c.}]\hat{\Phi}\}~,\lb{C8}\\
&& \hat{H}(\hat{\Phi})=H[\hat{B}(\hat{\Phi})]-\hat{\Phi}\hat{B}
(\hat{\Phi})\lb{C9}
\eea
 The linear terms with $C$ and $\bar{C}$ break the  $U_R(1)$-symmetry ,
however, this action is invariant with respect to the isometry
transformation.

It should be underlined that the $LG$-superfield $\hat{\Phi}$
transforms homogeneously, while the supersymmetry transformation of the
$LG$-chiral Lagrange multiplier $\phi$ contains the inhomogeneous term
\be
\delta_\epsilon\phi=
-i\xi(\theta^\alpha\bar{\epsilon}_\alpha)+i\epsilon^\alpha_k Q_\alpha^k
\phi~.
\lb{C10}
\ee

Consider the $\theta$-decomposition of the $LG$-chiral superfield
\be
\phi=A(x_{\s L}) +\theta^\alpha \psi_\alpha(x_{\s L})+(\theta)^2
F(x_{\s L})~,\lb{C11}
\ee
where  $x_{\s L}$ is the  coordinate of the chiral basis.

The Lie bracket of the modified supersymmetry transformation \p{C10}
 contains the composite central charge parameter corresponding to the
 following action of the generator $Z$ on the chiral superfield:
\be
Z \phi=\xi~,\qquad ( Z\bar{\phi} =-\xi )
~.\lb{C12}
\ee
Thus, the Goldstone boson field $\mbox{Im}\,A(x)$ for the central-charge
transformation appears in this model. It should be remarked that the
isometry transformation  in the chiral model without $PSBGS$ cannot be
identified with the central charge.

The vacuum  equations of motion for this model have the following form:
\bea
&& \bar{F}\hat{\tau}(b) + \bar{C}=0~,\qquad b=A+\bar{A}\lb{C13}\\
&& (|F|^2-\xi^2)\hat{\tau}^\prime(b)=0~,\qquad
  \hat{\tau}=\hat{H}^{\prime\prime}=-\tau^{-1}~.\lb{C15}
\eea

The scalar potential of this model depends on the one real scalar
component only
\be
{\cal V}(b)={1\over2}[\xi^2\hat{\tau}(b)+|C|^2\hat{\tau}^{-1}(b)]~.
\lb{C16}
\ee

The vacuum solution $|F_0|^2=\xi^2$ corresponds to the degeneracy
condition for the matrix of vacuum supersymmetry transformations. The
choice $F_0=i\xi$ breaks the 2-nd supersymmetry.

Thus, the non-trivial interaction of the $LG$-chiral superfield
$\phi$ provides the partial spontaneous breaking of the $D{=}3,~N{=}2$
supersymmetry. This phenomenon has been analysed also in the formalism
of the $D{=}3,~N{=}1$ Goldstone-type superfields \cite{IK}.

Let us assume that the spinor coordinates $\tao$ parameterize $N{=}1$
superspace, and the generators $Q_\alpha^1$ form the corresponding
subalgebra of the $N{=}2$ supersymmetry.

The chirality condition  in the real basis
\be
(\Dao+i\Dat )\phi=0
\lb{F4}
\ee
 can be solved via the complex unrestricted $N{=}1$ superfield
$\chi$
\be
\phi=\chi(x,\theta_1) +i\tat\Dao \chi(x,\theta_1) +\ttt \Doo
\chi(x,\theta_1)~.
\lb{F5}
\ee

The transformation \p{C10} generates the corresponding transformation of
the complex $N{=}1$ superfield:
\be
\delta\chi=-{1\over2}\xi\tao(\epsilon_{2\alpha}+i\epsilon_{1\alpha})
-i\epsilon^\alpha_2\Dao \chi+
i\epsilon^\alpha_1Q^1_\alpha \chi~.\lb{F8}
\ee

Consider the $\theta_2$-decomposition of the basic superfield \p{C5}
of the chiral $PSBGS$ model
\bea
&&\hat{\Phi} =\chi+\bar{\chi}+i\tat\Dao(\chi-\bar{\chi})+
\ttt\Doo(\chi+\bar{\chi})+i\xi[\too+\ttt]\nn\\
&&=\Sigma+\tat\Dao \rho+
\ttt[\Doo\Sigma+2i\xi]~,\lb{F10}
\eea
where $\Sigma$ is the massive real $N{=}1$ superfield and $\rho$ is
the real Goldstone superfield for the 2-nd supersymmetry.

The  $N{=}1$ superfields with the analogous $N{=}2$ transformations have
been proposed in ref.\cite{IK}. The authors of this work have shown that
the additional superfield  can be constructed in terms of the spinor
derivative of the Goldstone superfield $\rho$ in order to built the
supermembrane action. In our approach, the massive degrees of freedom
can be removed using the covariant condition
\be
\hat{\Phi}^2=0~,\lb{F16b}
\ee
which allows us to construct $\Sigma$ via $\Dao \rho$ by analogy with
the similar construction in the $D{=}4,~N{=}2$ theory \cite{RT}.

Let us analyse the $N{=}1$ decomposition of the gauge prepotential
\be
V_s(x,\theta_1,\theta_2)=\kappa(x,\theta_1)+i\tat V_\alpha(x,\theta_1)+
i\ttt M(x,\theta_1)\lb{F17}
\ee
and the chiral gauge parameter $\Lambda =\mbox{exp}(i\tat\Dao)
\lambda(x,\theta_1)$.

The gauge transformations of the $N{=}1$ components are
\be
\delta_\lambda \kappa=\lambda+\bar{\lambda}~,\qquad
 \delta_\lambda V_\alpha=\Dao(\lambda-\bar{\lambda})~.\lb{F20}
\ee
Thus, $\kappa$ is a pure gauge degree of freedom, $V_\alpha$ is the
$N{=}1$ gauge superfield, and $M$ is the scalar $N{=}1$  component of
the $N{=}2$ supermultiplet.

Consider the $N{=}1$ decomposition of the linear superfield \p{B39}
\be
W_s(V_s)
=w +i\tat F_\alpha(V)-\ttt\Doo w~,\lb{F25}
\ee
where  the gauge-invariant scalar and spinor superfields are defined
\bea
&&w={1\over2}[M+i\Doo\kappa]~,\lb{F26}\\
&& F_\alpha(V) ={i\over2}\Doo V_\alpha+{1\over4}\pab V^\beta~,\qquad
D^\alpha F_\alpha =0~. \lb{F27}
\eea

The Goldstone transformation of $W_s$ \p{B40} produces the
$\epsilon_2$-transformations of the $N{=}1$ superfields. The spinor
superfield strength $F_\alpha$ is analogous to the Goldstone spinor
superfield of ref.\cite{IK}. It describes the Goldstone degree of freedom
of the $D2$-brane, and the superfield $w$ corresponds to the massive
degrees of freedom. Our construction introduces the $N{=}1$ gauge
superfield $V_\alpha$ as the basic object of this model and allows us to
study the modification of the supersymmetry algebra on the gauge fields of
the $D2$-brane.

The author is grateful to  E.A. Ivanov, S.O. Krivonos, A.A. Kapustnikov,
O. Lechtenfeld and, especially, J. Lukierski for stimulating discussions
and thanks J. Lukierski and Z. Popowicz for the kind hospitality
in Wroclaw University where the main part of this work has been made.
The work  is partially supported by the grants RFBR-99-02-18417,
INTAS-93-127-ext, INTAS-96-0308, by the Bogoliubov-Infeld programme and by
the Uzbek Foundation of Basic Research, contract N 11/27.

\end{document}